%% file: main.tex
\documentclass[preprint]{elsarticle}
\usepackage{lmodern}
\usepackage{units}
\usepackage{epstopdf}
\usepackage{graphicx}
\usepackage{wrapfig}
\usepackage{multicol}
\usepackage{physics}
\usepackage{amsmath}
\usepackage{mathtools}
\usepackage{tikz}
\usepackage{mathdots}
\usepackage{yhmath}
\usepackage{cancel}
\usepackage{color}
\usepackage{array}
\usepackage{multirow}
\usepackage{amssymb}
\usepackage{textcomp}
\usepackage{gensymb}
\usepackage{tabularx}
\usepackage{extarrows}
\usepackage{booktabs}
\usepackage[font=footnotesize,labelsep=space]{caption}
\usepackage{subcaption}
\usetikzlibrary{fadings}
\usetikzlibrary{patterns}
\usetikzlibrary{shadows.blur}
\usetikzlibrary{shapes}
\usepackage{pdfpages}
\usepackage{epstopdf}
\DeclareMathAlphabet\mathbfcal{OMS}{cmsy}{b}{n}
\usepackage{hyperref}
\hypersetup{nolinks=true}
\usepackage[capitalise]{cleveref}
\usepackage{glossaries}
\crefname{equation}{Eq.}{Eqs.}
\crefname{table}{Tab.}{Tabs.}
\usepackage{natbib}
\bibpunct[, ]{(}{)}{;}{a}{}{,}
\usepackage[export]{adjustbox}
\usepackage{float}
\usepackage{bm}
\usepackage{soul}
\usepackage{microtype} 
\usepackage{trackchanges}
\usepackage[colorinlistoftodos,prependcaption]{todonotes}
\input{input/macros}

\input{input/glossary}
\makeatletter
\def\ps@pprintTitle{%
  \let\@oddhead\@empty
  \let\@evenhead\@empty
  \def\@oddfoot{\reset@font\hfil\thepage\hfil}
  \let\@evenfoot\@oddfoot
}
\makeatother

\begin{document}
\title{Exploring the Influence of Parametrized Pulsatility on Left Ventricular Washout under LVAD Support: A Computational Study Using Reduced-Order Models}

\author[1]{M.~R.~Schuster\corref{cor1}}%
\ead{schuster@cats.rwth-aachen.de}
\author[1]{N.~Dirkes}
\author[2]{F.~Key }
\author[2]{S.~Elgeti}
\author[1]{M.~Behr}
\cortext[cor1]{Corresponding author}
\affiliation[1]{organization={Chair for Computational Analysis of Technical Systems, RWTH Aachen University},
country={Germany}}
\affiliation[2]{organization={Institute of Lightweight Design and Structural Biomechanics, TU Wien},
country={Austria}}

\begin{abstract} 
Introducing pulsatility in LVADs is known to reduce complications such as stagnation and thrombosis, but it is an ongoing topic of research what the optimal form is.  We present a framework consisting of parametrized full-order simulations, reduced-order models, and sensitivity analysis to systematically quantify the effects of parametrized pulsatility on washout.As a sample problem, we study the washout in an idealized 2D left ventricle and a parametrized sinusoidal LVAD flow rate. The framework yields speed-ups proportional to the number of samples required in the sensitivity analysis. In our setting, we find that short, intense pulses wash out the left ventricle best, while the time between consecutive pulses does not play a significant role.  

\end{abstract}

\maketitle
\input{input/introduction-motivation}
\input{input/MathematicalModeling}

\input{input/FOM-FE}

\input{input/ROM}
\input{input/SensitivityAnalysis}
\input{input/Stagnation_ROMS}
\input{input/Results}

\input{input/Discussion}

\input{input/Conclusion}
\bibliographystyle{tfcse}
\bibliography{main} 
\end{document}

%% file: input/macros.tex
\newcommand{\thickhline}{%
    \noalign {\ifnum 0=`}\fi \hrule height 1pt
    \futurelet \reserved@a \@xhline
}
\newcolumntype{"}{@{\hskip\tabcolsep\vrule width 1pt\hskip\tabcolsep}}

\newcommand{\pulsePause}{$\Delta t_{p}$}
\newcommand{\pulsePeriod}{$\Delta t_{f}$}
\newcommand{\washout}{$\Delta c$}

%% file: input/glossary.tex
\newacronym{annlabel}{ANN}{artificial neural network}
\newacronym{fnnlabel}{FNN}{feedforward neural network}
\newacronym{clabel}{$c$}{concentration}
\newacronym{dclabel}{DC}{discontinuity capturing}
\newacronym{doflabel}{DOF}{degrees of freedom}
\newacronym{glslabel}{GLS}{Galerkin$/$Least-squares}
\newacronym{femlabel}{FEM}{finite element method}
\newacronym{gprlabel}{GPR}{Gaussian process regression}
\newacronym{inslabel}{INS}{incompressible Navier-Stokes}
\newacronym{ibvplabel}{IBVP}{initial boundary value problem}
\newacronym{morlabel}{MOR}{model order reduction}
\newacronym{osilabel}{OSI}{oscillatory shear index}
\newacronym{podlabel}{POD}{proper orthogonal decomposition}
\newacronym{rbflabel}{RBF}{radial basis function}
\newacronym{romlabel}{ROM}{reduced-order model}
\newacronym{fomlabel}{FOM}{full-order model}
\newacronym{romslabel}{ROMs}{reduced order models}
\newacronym{vimlabel}{VIM}{virtual ink method}
\newacronym{pulseperiodlabel}{\pulsePeriod}{pulse period}
\newacronym{pulsepauselabel}{\pulsePause}{time between two subsequent pulses}
\newacronym{cfdlabel}{CFD}{computational fluid dynamics}
\newacronym{lvadslabel}{LVADs}{left ventricular assist devices}
\newacronym{lvadlabel}{LVAD}{left ventricular assist device}
\newacronym{1plabel}{$1P$}{single-pulse case}
\newacronym{2plabel}{$2P$}{two-pulse case}
\newacronym{lvlabel}{LV}{left ventricle}
\newacronym{snirom}{sniROM}{standardized non-intrusive reduced order modeling}
\newacronym{avlabel}{AV}{aortic valve}
\newacronym{mvlabel}{MV}{mitral valve}

%% file: input/introduction-motivation.tex
\section{Introduction}
\glsresetall
 Cardiovascular diseases culminating in heart failure are among the most common causes of death in Western countries \citep{GBD2019,ferreira2019world}. \Glspl{lvadlabel} have been used as bridge-to-transplant, bridge-to-candidacy, and destination therapy for patients suffering from severe heart disease. Such a device is implanted in the left ventricle and pumps blood through a cannula inserted at the apex to the aorta. 
Since the invention of \glspl{lvadlabel} in the 1960s \citep{DEBAKEY19713}, they have undergone multiple generations of innovation and nowadays consist most often of a rotational pump with a constant flow rate. They are referred to as continuous-flow \glspl{lvadlabel} (cf-LVADs). Although this design brings several advantages like durability, higher patient survival \citep{rose2001long}, and size reduction, it comes with the loss of pulsatility when compared to early generation displacement pumps. 

A lack of pulsatility has been proven to increase the risk for thrombosis and stroke \citep{mehra2017}. It has been shown that thrombosis is caused by stagnation and stagnant zones. Stagnation is inversely related to the amount of blood that is washed out of the left ventricle in a given time period, which we call \textit{washout}. Stagnation zones have been identified in the left ventricle around the cannula and close to the aortic valve as well as in the \gls{lvadlabel}, for an overview see \cite{reider2017}. This emphasizes that not only an estimate about the global stagnation in the left ventricle is needed, but also the location of stagnant areas is of high interest. 
Thresholds on velocity and shear rate are popular hemodynamic parameters to quantify stagnation \citep{liao2018, liao2016, wong2014}. A velocity-based stagnation index has been introduced in \cite{quaini2011}, whereas a pulsatility index is used elsewhere \citep{reider2017, liao2018}. Wall shear stress and quantities like \gls{osilabel} have been used as well \citep{zhang2017}. A full model for thrombus growth is introduced by \cite{menichini2016mathematical}. To track blood transport at each discrete location over time, approaches like residence time \citep{rossini2016} and the virtual ink method \citep{Rayz2009} have been introduced, where an additional transport equation is solved.
The topic of stagnation in the LVAD assisted heart has been investigated by experiments and numerical simulations.
An experimental study of the \gls{lvadlabel} speed modulation is presented by \cite{ZimpferLavare}, another experimental study finds a link between the formation of thrombi and stagnation zones in the left ventricle \citep{MayNewman}.
\Gls{cfdlabel} has been used for in-silico experiments to simulate different aspects of cardiovascular flows, for example, the cardiac function \citep{dede2021computational,verzicco2022electro,mittal2016computational,meschini2018flow}, hemolysis in LVADs \citep{hassler2019,wiegmann2019}, pump washout, stagnation, or platelet activation in \glspl{lvadlabel} \citep{fang2022,boraschi2021}, and blood flow in large vessels \citep{updegrove2017simvascular}. 
Simulations of the left ventricle with LVAD have been performed with fluid-structure interaction in \cite{bakir2018} and \cite{mccormick2013simulating}.
\Gls{cfdlabel} analysis allows to quantify stagnation zones and the connected thrombosis risk. 

The formation of thrombus in the left ventricle and LVAD is a multifaceted phenomenon that has been studied extensively. Studies have been performed on parameters such as the shape of the cannula tip \citep{wong2014,liao2016}, the length of the inserted cannula \citep{chivukula2020}, the positioning of the cannula in the left ventricle \citep{prisco2017,ghodrati2020}, and the angle of cannula insertion \citep{schloeglhofer2022,neidlin2021understanding}. Adding pulsatility to the flow rate of the \gls{lvadlabel} may reduce the risk of thrombosis and stroke in the left ventricle and in the LVAD itself \citep{ZimpferLavare, ILSJarvik}. LVAD speed modulations have been used to restore pulsatility, such as the Lavare cycle or the Heartmate3 artificial pulse \citep{fang2022}. The synchronicity of such pulses with native left ventricular activity is an ongoing topic of research \citep{liao2018,maynewman2023,mccormick2014}.

The complex relationship between LVAD implantation and thrombosis risk is a problem with many parameters and their influence on stenting is not straightforward. It is of interest to quantify how sensitive washout is to these parameters. Performing such a sensitivity analysis requires many data points, which are expensive to generate from full-order numerical simulations. A reduction in simulation cost is desirable. To increase the practicality of CFD simulations and make them relevant to clinical practice, it is necessary to reduce computational cost while maintaining high fidelity. One way to achieve faster results is to use simplified models, as in \cite{moulton2017simulation} for the structural model of the muscle fiber. Another way to reduce simulation time is to approximate \glspl{fomlabel} with \glspl{romlabel}. For cardiovascular simulation, such models have recently become popular \citep{karabelas2022global,collia2022surrogate,di2019reduced,di2020model}. 

Therefore, the aim of this study is to present a method to evaluate the sensitivity analysis using \glspl{romlabel} in order to reduce the computational cost. As a model problem for such an analysis, we chose an idealized, simplified 2D geometry of the left ventricle with an inserted LVAD cannula. 

In this model problem, we want to determine how sensitive washout in an idealized 2D left ventricle with inserted cannula is to using a parametrized pulsatility in the LVAD flow rate. The pulse is parametrized as a sinuisodal variation of a baseline flow rate, with the parameters being the length of the pulse, the amplitude of flow rate increase, and the time between two subsequent pulses serve as the parameters.

To decrease the computational cost, a standardized non-intrusive \gls{romlabel}, as introduced in \cite{berzins2021}, will be used here. It is based on full-order model snapshots created by \gls{cfdlabel} simulations with the \gls{femlabel}. We use a local concentration and an integral quantity based on the virtual ink method to compute the washout. For the full-order simulations, we use a simplified 2D domain representing the left ventricle and the LVAD cannula. We use \gls{gprlabel} and \gls{annlabel} to find the coefficients of the \gls{romlabel}. Eventually, we perform variance-based sensitivity analysis for the influence of each parameter on washout. The decreased computational cost allows for the evaluation of washout for a larger number of data points in the entire parameter space. This is needed for the sensivity analysis to find the underlying relation between parameters and washout. 

%% file: input/MathematicalModeling.tex
\section{Methods}
\label{sec:methods}
In the following section, we describe how we model blood flow in the left ventricle as an incompressible Newtonian fluid with the Navier-Stokes equations and washout based on the virtual ink method as a scalar advection equation. We outline how we discretize these equations using \gls{femlabel} and we show the methods we use to create the \gls{romlabel} and to perform sensitivity analysis.
\subsection{Mathematical Modeling of Flow and Washout in the Left Ventricle}
\label{sec:mathmodels}
\begin{figure}
	\begin{centering}
		\input{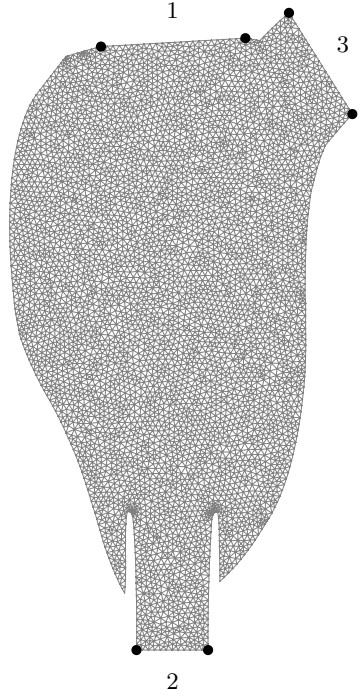}
		\captionsetup{justification=raggedright,
			singlelinecheck=false,width=\textwidth}
		\caption{The domain and computational grid for the 2D simulations of the blood flow in the left ventricle with the boundaries (1) mitral valve ($=\Gamma_{\textrm{in}})$, (2) LVAD cannula ($=\Gamma_{\textrm{out}})$, and (3) aortic valve together with the rest of the boundary of the left ventricle($=\Gamma_{\textrm{w}}$) . The black dots mark the respective boundary ends. }
		\label{fig:comp_domain}
	\end{centering}
\end{figure}

In the healthy left ventricle, blood enters from the atrium through the mitral valve during diastole and exits into the aorta through the aortic valve, whereas in the assisted left ventricle, the LVAD is attached to the apex and bypasses the aortic valve directly into the aorta. Blood still enters through the mitral valve. Because of the reduced pressure in the ventricle, the aortic valve remains closed most of the time. Blood flow is driven by the LVAD pump, which extracts blood at a flow rate determined by its rotational speed. 
We use an idealized, non-deforming, 2D left ventricular geometry; see \cref{fig:comp_domain}. The 2D geometry is obtained from a planar slice through a 3D geometry obtained from a numerical simulation \citep{pfaller2019}. We choose the largest volume of this simulated heart beat as our base geometry. The largest volume is chosen because the original geometry does not simulate a diseased patient, but should account for the increased left ventricular volume in LVAD patients. The LVAD cannula is inserted at the apex. The plane that makes up our 2D geometry is defined by the center of the aortic valve, mitral valve, and LVAD cannula, as shown in \cref{fig:comp_domain}. The valves are defined as line segments. The domain boundaries consist of the aortic valve, mitral valve, cannula outlet and the inner ventricular walls. 

\begin{figure}
	\begin{center}
		\captionsetup{width=1\textwidth}
		\includegraphics[width=0.8\textwidth]{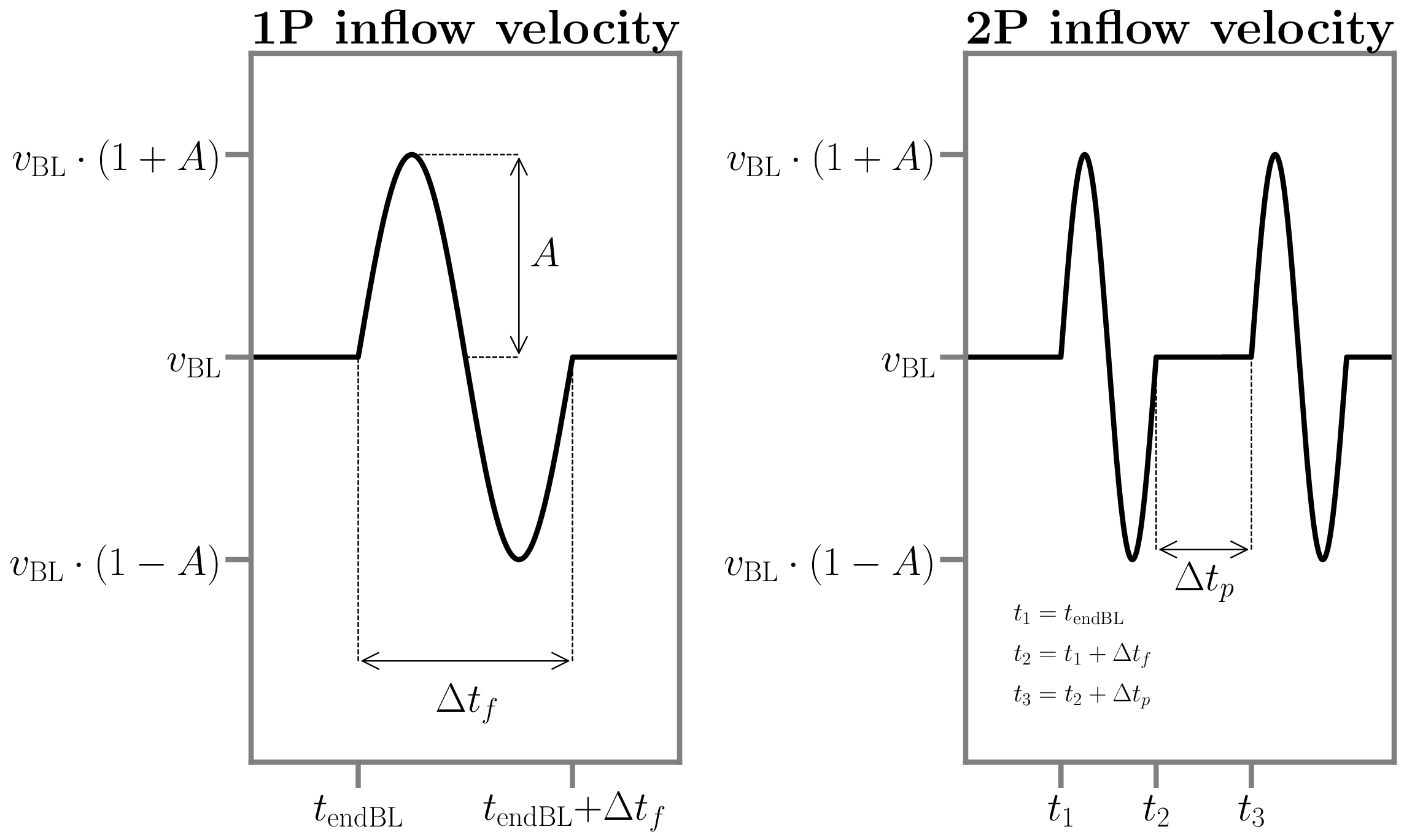}
		\caption{Parametrization of the LVAD flow rate as a velocity boundary condition with 3 parameters: \pulsePeriod{}, \pulsePause{} and $A$. The baseline velocity is labeled $v_{\mathrm{BL}}$.\label{fig:fluxdisc}}
	\end{center}
\end{figure}

The LVAD flow rate is modeled as a time-dependent velocity expression and prescribed at the mitral valve. The prescribed velocity profile is parabolic over boundary width. At the outlet, a traction-free outflow is prescribed. The aortic valve remains closed at all times. At the inner surface of the left ventricular wall, the velocity is zero. 

For the pulsatility of the \gls{lvadlabel}, we have developed a parametrized flow rate curve. It is modeled as a baseline flow equivalent to 5 $\mathrm{L/min}$ with an additional sinusoidal variation in time. Note that in the 2D case, the inlet is a 1D line, whereas the equivalent of 5 $\mathrm{L/min}$ has been calculated assuming a 2D circular inlet with parabolic inflow profile. 

We investigate two scenarios for the parametrized inflow velocity: a single pulse (\gls{1plabel}), where the vector of parameters $\boldsymbol{\mu}$ contains the amplitude $A$ between highest flow rate and baseline, and the length of the pulse \pulsePeriod{}, $	\boldsymbol{\mu}_{1P} = [A, \Delta t_f]$; and two subsequent pulses (\gls{2plabel}), where the variable parameters are the time between two subsequent pulses \pulsePause{} and \pulsePeriod{} with a fixed $A$, $	\boldsymbol{\mu}_{2P} = [ \Delta t_f, \Delta t_p]$; see \cref{fig:fluxdisc}.

We split the boundaries into the inlet $\Gamma_{\textrm{in}}$, the outlet $\Gamma_{\textrm{out}}$, and the wall boundaries $\Gamma_{\textrm{w}}$ and use the incompressible Navier-Stokes equations with velocity $\mathbf{u} = \mathbf{u} (\boldsymbol{x},t,\boldsymbol{\mu})$ and pressure $p = p (\boldsymbol{x},t,\boldsymbol{\mu}) $ to model laminar blood flow in the steady domain $\Omega$ as:

\begin{subequations}
	\label{eq:nse}
	\begin{alignat}{2}
		\boldsymbol{\nabla} \cdot \mathbf{u} &= 0 &&\qquad~\mathrm{in} ~\Omega,\\
		{\rho} \left( \frac{\partial \mathbf{u}}{\partial t} + \mathbf{u} \cdot  \boldsymbol{\nabla}  \mathbf{u} \right) - \boldsymbol{\nabla} \cdot \boldsymbol{\sigma} & = 0  &&\qquad~\mathrm{in} ~\Omega,\\
				\mathbf{u} & = \mathbf{u} (\boldsymbol{x},t, \boldsymbol{\mu}) &&\qquad~\mathrm{on} ~\Gamma_{\textrm{in}}, \\
		\mathbf{u} & = \mathbf{0} &&\qquad~\mathrm{on} ~\Gamma_{\textrm{w}}, \\
		\boldsymbol{\sigma} \cdot \mathbf{n} & = \mathbf{h} &&\qquad~\mathrm{on}~ \Gamma_{\textrm{out}}, \\
		\boldsymbol{u}(t=0) & = \boldsymbol{u_0} &&\qquad~\mathrm{in} ~\Omega.
	\end{alignat}
\end{subequations}

The fluid model is closed by the relation between stress and strain via the Cauchy stress tensor $\boldsymbol{\sigma}$, with the identity tensor $\mathbf{I}$ and the fluid viscosity $\mu$:
\begin{subequations}
\begin{align}
\boldsymbol{\sigma} = - p \mathbf{I} + 2 \mu \boldsymbol{\epsilon}, \qquad\\
\boldsymbol{\epsilon} = \frac{1}{2} \left( \boldsymbol{\nabla} \mathbf{u} + \boldsymbol{\nabla} \mathbf{u}^T \right).
\end{align}
\end{subequations}

Next, we introduce the measure for washout based on the virtual ink method \citep{Rayz2009}. The transport of blood is modeled as a passive tracer being convected by the fluid velocity $\mathbf{u}$. For our adaption of the method, the initial boundary value problem of a scalar advection equation for the marker concentration $c = c(\boldsymbol{x}, t, \boldsymbol{\mu})$ is solved:
\begin{subequations}
	\label{eq:ST}
	\begin{alignat}{2} 
		\frac{\partial c}{\partial t} + \mathbf{u} \cdot  \boldsymbol{\nabla}  c & =0  &&\qquad~\mathrm{in} ~\Omega,\\
		c(\mathbf{x},0) & = c_0 = 1 &&\qquad~\mathrm{in} ~\Omega,\\
		c(\mathbf{x},t) & = c_D =  0 &&\qquad~\mathrm{on} ~\Gamma_{\textrm{in}},\\
		\boldsymbol{\nabla} c \cdot \mathbf{n} & = 0 &&\qquad~\mathrm{on}~ \Gamma_{\textrm{w}} \cup \Gamma_{\textrm{out}} .
	\end{alignat}
\end{subequations}

 We calculate the washout \washout{} by evaluating how much of the \textit{initially marked blood} remains in the left ventricle after a certain period of time. The \textit{initially marked blood} is labeled with an initial concentration value $c_0=1$. \textit{Newly marked blood} entering through the mitral valve is assigned a different concentration value $c_{\textrm{D}}=0$ at the mitral valve. A zero flux condition ($\boldsymbol{\nabla} c \cdot \mathbf{n} = 0)$ is imposed at the walls $\Gamma_{\textrm{w}}$. The evolution of the concentration is tracked by solving \cref{eq:ST}. The washout \washout{} is computed as the difference between the integrated initial concentration field at $t_0$ and the integrated concentration field at the final time step $t_{\textrm{end}}$:
\begin{align}
	\Delta c ( \boldsymbol{\mu}) = 1- \frac{\int_{\Omega} c (\boldsymbol{x},t_{\mathrm{end}}, \boldsymbol{\mu})~ \mathrm{d}V }{\int_{\Omega} c \left(\boldsymbol{x},t_0\right) ~\mathrm{d} V_{t_{0}}}=1- \frac{\int_{\Omega} c (\boldsymbol{x},t_{\mathrm{end}},\boldsymbol{\mu})~ \mathrm{d}V }{\int_{\Omega} 1 ~\mathrm{d} V}.\label{eq:washout}
\end{align}

For simplicity of computing \washout{}, we have replaced the integral over the domain by the sum of concentration values at all nodes of the mesh, divided by the total number of nodes $n_\mathrm{n}$.

%% file: input/FOM-FE.tex
\subsection{Full-Order Model: Finite Element Discretization}
\label{sec:fom-fe}

\Cref{eq:nse} and \cref{eq:ST} are solved with a stabilized Galerkin finite element method with a time-discontinous prismatic space-time formulation \citep{Pauli2017, danwitz2023}. Considering the variational problem related to \Cref{eq:ST}, a modified formulation with YZ$\beta$ discontinuity capturing \citep{Bazilevs2007} is used, with the diffusion term $\nu_{\textsf{\tiny{DC}}} = \nu_{\textsf{\tiny{DC}}}(Y, Z, \beta)  $. $Y$ is a reference concentration value, which we set to be $c_{\textsf{\tiny{ref}}}  = 1 $, $Z$ is the residual of the transport equation \cref{eq:ST}, and $\beta$ is a free parameter that ranges from 1 for smooth layers to 2 for sharp layers, we use $\beta=1.4$. All simulations are carried out with an MPI parallelized in-house solver on the RWTH Claix 2018 compute cluster.

%% file: input/ROM.tex
\subsection{Reduced-Order Model for Washout}
\label{sec:ROM}
In this section, we illustrate the steps to create a reduced-order model (ROM) for the concentration values based on the virtual ink method of the full-order simulation. The method employed here has been presented as standardized non-intrusive reduced-order modeling in \cite{berzins2021}. It is available as open source\footnote{https://github.com/arturs-berzins/sniROM}. We have modified the original source code to work with the output of the scalar advection equation (\Cref{eq:ST}). 

The FOM of the VIM is parametrized through the inflow velocity profile in \Cref{eq:nse}. The local concentration values $c$ are functions of the spatial coordinates $\boldsymbol{x}$ and of the parameters $\boldsymbol{\mu}$.  The washout \washout{}, (see \Cref{eq:washout}), of the \gls{1plabel} and \gls{2plabel} scenarios is a function of the parameters $\boldsymbol{\mu}$ (see \Cref{fig:fluxdisc}) :
\begin{subequations}
	\label{eq:parametrization}
	\begin{alignat}{1}
	c_{1P} &=c(\boldsymbol{x},\Delta t_f, A),\\
	\Delta c_{1P} &= \Delta c (\Delta t_f,A),\\
	c_{2P} &=c(\boldsymbol{x},\Delta t_p, \Delta t_f),\\
	\Delta c_{2P} &= \Delta c(\Delta t_p, \Delta t_f).
	\end{alignat}
\end{subequations}

The simulation data is obtained from a training set of samples of the respective parameter spaces of the two scenarios, see \Cref{tab:sim-params}.

The main steps to create the \gls{romlabel} are: $(i)$ assemble the virtual ink method simulation data in a snapshot matrix, $(ii)$ perform \gls{podlabel} of the snapshot matrix to find the \textit{reduced basis functions} and \textit{reduced coefficients}, and $(iii)$ find a mapping between the input parameters and the simulation data by means of a regression on the \textit{reduced coefficients}. Whenever details are skipped or only briefly mentioned, the reader can refer to \cite{berzins2021}.

\subsubsection{Reduced Basis and Reduced Coefficients}

The solution of the parametrized full-order problem is the concentration value ${c} (\boldsymbol{x},\boldsymbol{\mu})$. From this solution, the quantity of interest \washout{} is computed per \cref{eq:washout}. It can be expressed with the basis functions $\boldsymbol{\phi} (\boldsymbol{x})  = \left[ \phi_1 (\boldsymbol{x}) | \dots | \phi_{N_{h}} (\boldsymbol{x}) \right] ^{\textsf{T}} \in \mathbb{R}^{N_{h}} $  and parameter-dependent coefficients $\boldsymbol{\alpha}^{\textsf{T}} (\boldsymbol{\mu}) \in \mathbb{R}^{N_\textrm{h}}$, where $N_\textrm{h} = n_{\text{n}} \cdot 1$ is the size of the FOM system, (for the solution, we only consider the nodes at the upper space-time level $n_{\textrm{n}}$ of the final time step, such that the number of degrees of freedom $n_{\mathrm{dof}}=1$):
\begin{align}
	c (\boldsymbol{x}; \boldsymbol{\mu}) = \boldsymbol{\alpha}^{\textsf{T}} (\boldsymbol{\mu}) \boldsymbol{\phi} (\boldsymbol{x}).
\end{align}

To reduce the dimensionality compared to the FOM, we seek a ROM that finds an approximation ${\boldsymbol{\alpha}}_L (\boldsymbol{\mu}) \in \mathbb{R}^{N_h}$ for the solution coefficient vector $\boldsymbol{\alpha} (\boldsymbol{\mu})$ by a linear combination of a reduced basis  {\boldmath $\mathsf{V}$}$=\left[ \boldsymbol{v}^{(1)} | \dots | \boldsymbol{v}^{(L)} \right] \in \mathbb{R}^{N_\mathrm{h} \times L}$ of $L$ basis vectors, with $L \ll N_h$, and the reduced coefficients  $\boldsymbol{y}(\boldsymbol{\mu}) = \left[ y_1 | \dots | y_L \right]^\textsf{T} \in \mathbb{R}^L$:
\begin{align}
	\boldsymbol{\alpha} \left(\boldsymbol{\mu} \right)  \approx \boldsymbol{\alpha}_L (\boldsymbol{\mu})= \sum_{l=1}^L y_l(\boldsymbol{\mu}) \boldsymbol{v}^{(l)}(\boldsymbol{x} ) = \bm{\mathsf{V}} \boldsymbol{y}(\boldsymbol{\mu}).
\end{align}
The reduced-order solution $\hat{c}$ can be reconstructed in the full space with the basis functions $\boldsymbol{\phi}(\boldsymbol{x})$ of the full-order as:\\
\begin{align}
 \hat{c} \left(\boldsymbol{x};\boldsymbol{\mu} \right) = \boldsymbol{y}^{\mathsf{T}}(\boldsymbol{\mu}) \bm{\mathsf{V}}^{\mathsf{T}}\boldsymbol{\phi} \left(\boldsymbol{x} \right). \label{eq:ROM-RB}
\end{align} 
\subsubsection{Proper Orthogonal Decomposition of the Snapshot Matrix}
\label{sec:pod}
Next, we will explain how to construct the reduced basis $\boldsymbol{\mathsf{V}}$ by \gls{podlabel} using the snapshot matrix. The snapshot matrix is assembled column-wise by $N_{\textrm{tr}}$ solution vectors of the virtual ink method full-order solution, where $N_{\textrm{tr}}$ is the number of training data samples. $\bm{\mathsf{S}} = [ \boldsymbol{\alpha}(\boldsymbol{\mu}^{(1)} ) | \dots | \boldsymbol{\alpha}(\boldsymbol{\mu}^{(N_\mathrm{tr}} ) ] \in \mathbb{R}^{N_h \times N_{\mathrm{tr}} } $ is decomposed into two orthonormal matrices $\bm{\mathsf{W}} \in \mathbb{R}^{N_h \times N_h}$ and $\bm{\mathsf{Z}} \in \mathbb{R}^{N_{\mathrm{tr}} \times N_{\mathrm{tr}}}$ containing the left and right singular vectors, and the diagonal matrix $\bm{\mathsf{\Sigma}} \in \mathbb{R}^{N_{h} \times N_{\mathrm{tr}}}$, containing the singular values in descending order:
	\begin{align}
			\frac{\bm{\mathsf{S}}}  { \sqrt{N_{\mathrm{tr}}}} = \bm{\mathsf{W}} \bm{\mathsf{\Sigma}} \bm{\mathsf{Z}} ^{\mathsf{T}}.
	\end{align} 
\gls{podlabel} creates an orthonormal basis $\bm{\mathsf{V}}$, such that $\bm{\mathsf{V}} \bm{\mathsf{V}} ^{\mathsf{T}} = \bm{\mathsf{I}}$. Here, $\bm{\mathsf{V}}$ is made up of the first $L$ left singular vectors of the matrix $\bm{\mathsf{W}}$. With the orthonormal basis, we can obtain obtain the reduced coefficients by projection of the \gls{fomlabel} solutions onto the reduced basis:
\begin{align}
	\boldsymbol{y} (\boldsymbol{\mu}) = \bm{\mathsf{V}}^{\mathsf{T}} \boldsymbol{\alpha}(\boldsymbol{\mu}).
\end{align}

\subsubsection{Standardization}
We perform snapshot centering, parameter standardization, and reduced coefficient standardization. Note that quantities indexed with \textit{c} are centered, and quantities indexed with \textit{s} are standardized.

The centered snapshot matrix $\bm{\mathsf{S}}^c = [ \boldsymbol{\alpha}^c (\boldsymbol{\mu}^{(1)}) | \dots | \boldsymbol{\alpha}^c (\boldsymbol{\mu}^{(N_{tr})}) ] $ is obtained by subtraction of the mean over all columns $\boldsymbol{\bar{\alpha}}$ per snapshot:
\begin{align}
	\boldsymbol{\alpha}^c (\boldsymbol{\mu}) = 	\boldsymbol{\alpha} (\boldsymbol{\mu}) - \bar{\boldsymbol{\alpha}}.
\end{align}

A major advantage of centering is that the reduced coefficients are zero mean, an assumption also made by the Gaussian process. This applies to the centered reduced basis $\bm{\mathsf{V}}^c$, which ensures that the centered reduced coefficients $ \boldsymbol{y}^c (\boldsymbol{\mu}) = \bm{\mathsf{V}}^c \boldsymbol{\alpha}^c (\boldsymbol{\mu})$ are also zero mean.
The standardized reduced coefficients are obtained by scaling the centered reduced coefficients by their standard deviations:
\begin{align}
	\boldsymbol{y}^s (\boldsymbol{\mu})=  (\bm{\mathsf{\Sigma}}^c)^{-1} \boldsymbol{y}^c (\boldsymbol{\mu}).
\end{align} 
Finally, the parameters $\boldsymbol{\mu}$ are standardized by their mean $\bar{\boldsymbol{\mu}} = 1/ N_{tr} \sum_{ \boldsymbol{\mu} \in \mathbb{P}_{tr}} \boldsymbol{\mu}$ and their standard deviation  $\bar{\bar{\boldsymbol{\mu}}} =  \sqrt{1/ N_{tr} \sum_{ \boldsymbol{\mu} \in \mathbb{P}_{tr}} (\boldsymbol{\mu}-\bar{\boldsymbol{\mu}})^2}$:
\begin{align}
	\boldsymbol{\mu}^s = (\boldsymbol{\mu} - \bar{\boldsymbol{\mu}} ) / \bar{\bar{\boldsymbol{\mu}}}.
\end{align}

\subsubsection{Regression Models}	\label{sec:reg}
\glsreset{gprlabel} 
\glsreset{annlabel}
Finally, we need to compute  $\boldsymbol{y} (\tilde{\boldsymbol{\mu}})$  when $c$ is evaluated for any new parameter sample $\tilde{\boldsymbol{\mu}}$. 
A mapping  $\boldsymbol{\pi} : \boldsymbol{\mu} \mapsto \boldsymbol{y} (\boldsymbol{\mu})$ from the parameters $\boldsymbol{\mu}$ to the reduced coefficients $\boldsymbol{y}$ is provided by fitting a regression model, which connects the input of the set of parameter samples $\mathbb{P}_{tr}$ with the output of the respective reduced coefficents $\mathbb{Y}_{tr} = \{ \boldsymbol{y}(\boldsymbol{\mu})_{\boldsymbol{\mu} \in \mathbb{P}_{tr}} \}$. The regression model provides predicted reduced coefficients $\tilde{\boldsymbol{y}}$ for new parameter samples $\tilde{\boldsymbol{\mu}}$. By projection to the full space, a predicted solution $\tilde{\boldsymbol{\alpha}}_L$ can be obtained:
\begin{align}
	\tilde{\boldsymbol{\alpha}}_L (\tilde{\boldsymbol{\mu}})= \bm{\mathsf{V}} \boldsymbol{y} (\tilde{\boldsymbol{\mu}}) = \bm{\mathsf{V}} \boldsymbol{\pi}(\tilde{\boldsymbol{\mu}}). \label{eq:proj}
\end{align}
Then \cref{eq:ROM-RB} is used to create the approximated solution by interpolation with the FOM basis functions.

There are two regression models used in this study: {\Gls{gprlabel}} and \gls{annlabel}. \Gls{gprlabel} establishes a scalar regression function for each standardized reduced coefficient $\alpha_l^s$ of a single component $l \in [1,L]$:
	\begin{align}
		\tilde{\pi}_l^s : \mathbb{R}^{N_d} \to \mathbb{R}, ~~\pi_l^s (\boldsymbol{\mu}) = \alpha^s_l.
	\end{align}
There is no assumption made about any parametric form of the function, rather the probability distribution of all functions that match the observed data is found. With this distribution we can predict the expected output at some unseen input. A key assumption is that the output vector of the regression  $\boldsymbol{y}_l^s$ follows a multivariate Gaussian distribution:
	\begin{align}
		\boldsymbol{y}^s_l  \sim  \mathcal{N} (\boldsymbol{m}, \bm{\mathsf{{K}}}),
	\end{align} 
	with the mean  vector $\boldsymbol{m}$, the covariance matrix $\bm{\mathsf{{K}}}$ and the $N_{\mathrm{tr}}$ input samples $\bm{\mathsf{P}} = [ (\boldsymbol{\mu}^s)^{(1)} | ... | (\boldsymbol{\mu}^s)^{(N_{\mathrm{tr}})}] \in \mathbb{R}^{N_d \times N_{\mathrm{tr}}}$  :
	\begin{align}
	\bm{m} &= m (\bm{\mathsf{P}}) := [ m((\boldsymbol{\mu}^s)^{(i)})]_{  1 \leq i \leq N  } \in \mathbb{R}^{N_{\mathrm{tr}}} \\
	\bm{\mathsf{K}} &= k(\bm{\mathsf{P}},\bm{\mathsf{P}}) := [ k ( (\boldsymbol{\mu}^s)^{(i)}, (\boldsymbol{\mu}^s)^{(j)} )  ] _{  1 \leq i,j \leq N  } \in \mathbb{R}^{N_{\mathrm{tr}} \times N_{\mathrm{tr}}}.
		\end{align}
The continuous regression function will then be a Gaussian process (GP): 
\begin{align}
	\tilde{\pi}_l^s  \sim \mathcal{GP} (m(\boldsymbol{\mu}^s),k( \boldsymbol{\mu}^s,\boldsymbol{\mu}^s_{\star} )),
\end{align}
with the mean $m( \boldsymbol{\mu}^s)$ and all possible input pairs $( \boldsymbol{\mu}^s,\boldsymbol{\mu}^s_{\star} )$. The mean vector and covariance matrix can be understood as finite-dimensional manifestations of the infinite-dimensional counterparts of a Gaussian distribution.

The assumption for the mean function is a zero mean, $m \left(\boldsymbol{\mu} \right) = 0$.
Further assumptions are made for the covariance function. Most importantly, \textit{stationarity} of the inputs is used, stating that the covariance between two outputs is only a function of the Euclidean distance of them in the input space.
We use the \textit{anisotropic squared exponential kernel}, which includes the unique correlation length $d_i$, where $i$ is the input dimension, the prior covariance $\sigma_f$, and the additional observational noise $\epsilon$ to better deal with numerical issues:
\begin{align}
	k \left( \boldsymbol{\mu}^s, \boldsymbol{\mu}^s_{\star}; \sigma_f, \boldsymbol{d}, \epsilon \right) = \sigma^2_f ~\mathrm{exp} \left( - \sum_{i=1}^{N_d} \frac{\left( \mu_i^s - \mu^s_{\star i}\right)^2}{2d^2_i}  \right) + \delta, \mathrm{ ~where ~} \delta =
	\begin{cases}
				\epsilon ~ \text{if}~ \boldsymbol{\mu}^s = \boldsymbol{\mu}_{\star}^{s'},\\
				0 ~ \text{otherwise.}
			\end{cases}
\end{align}

The \textit{maximum likelihood estimation} is used to find the hyperparameters $ \mathbb{H} = \lbrace \sigma_f, \boldsymbol{d}, \epsilon \rbrace$. More details on \gls{gprlabel} can be found in \cite{berzins2021}.
Hyperparameter tuning for \gls{gprlabel} is performed on a validation set of 24 samples. All samples are drawn from the parameter space based on \textit{Latin Hypercube Sampling}.

The second regression model, ANN, is based on a feedforward neural network. Here, we will only shortly discuss the architecture of the preferred network. The loss function is the standardized regression error $\boldsymbol{\varepsilon}_{ANN}$, such that the optimization problem determines the weights $\Theta$:
\begin{align}
	\Theta = \underset{\hat{\Theta}}{\text{argmin}} \sqrt{\frac{1} { N_{tr}} \sum_{\boldsymbol{\mu} \in \mathbb{P}_{tr}} \boldsymbol{\varepsilon_{\textsf{ANN}}^2} \left( \boldsymbol{\mu}; \tilde{\boldsymbol{\pi}}_l^s \left( \boldsymbol{\mu}^s; \hat{\theta} \right) \right) } = \underset{\hat{\Theta}}{\text{argmin}}~ \varepsilon_{\textsf{ANN}} \left( \mathbb{P}_{tr}; \hat{\Theta} \right).
\end{align}

The training is performed using gradient-based iterative optimizers with backpropagation. We compute a weight update in each iteration $i$ with a mini-batch $\mathcal{P}_b \in \mathcal{P}_{\mathrm{tr}}$, with $N_b$ elements of the training set:
\begin{align}
	\Theta^{i+1} = \Theta^i - \alpha G \left( \frac{\partial \varepsilon_{\textsf{ANN}} \left( \mathbb{P}_b ; \Theta^i \right) }{\partial \Theta^i} \right).
\end{align}

Due to standardization, we use the same parameters as in \cite{berzins2021}, which are repeated here for completeness.
We use $N_b = 10$ with 5000 \textit{epochs}, equivalent to 5000 $N_{tr}/N_b$ iterations. $\alpha$ is the learning rate and $G$ depends on the specific optimizer used. We use the Adam optimizer with the Adam hyperparameter values $\epsilon = 10^{-8}, \beta_1 = 0.9, \beta_2 = 0.999$. The FNN is restricted to be a shallow ANN with two hidden layers $N^j = 2$ of $N_v$ neurons each \citep{hesthaven2018non}. We employ a hyperbolic tangent activation function $g(x) = \tanh(x)$. \textit{Hyperparameter tuning} is performed on $\alpha_0$. This is implemented by performing a grid-search over the hyperparameter space $\left(\alpha_0 \times N_v \right) \in \left[ 10^{-4},1 \right] \times \left[10,50 \right]$ with the Python Tune module \citep{liaw2018tune}. The best learning rate is supposed to be few orders of magnitude below the order of magnitude of the data due to standardization $\mathcal{O}(1)$. The number of learnable degrees of freedom should be close to the number of outputs in the training data. We use $L = 15$ and $N_{tr} = 96$, expecting $N_v$ to be around 35. For the implementation of the ANN we make use of PyTorch \citep{pytorch}.

\subsubsection{Errors}
The difference between a full-order solution coefficient vector $\boldsymbol{\alpha} (\boldsymbol{\mu})$ and its rank $L$ approximation $\boldsymbol{\alpha}_L (\boldsymbol{\mu})$ can be calculated as the Euclidean distance:
\begin{align}
	\delta_{\text{POD}} \left(\boldsymbol{\mu};\boldsymbol{\mathsf{V}} \right) = || \boldsymbol{\alpha}\left(\boldsymbol{\mu} \right) - \boldsymbol{\alpha}_L\left( \boldsymbol{\mu}  \right) ||_2 = || \boldsymbol{\alpha}\left(\boldsymbol{\mu}\right) - \boldsymbol{\mathsf{V}}\boldsymbol{\mathsf{V}}^T \boldsymbol{\alpha}\left( \boldsymbol{\mu}  \right)  ||_2.
	\label{eq:dpod}
\end{align}

The error introduced in \cref{eq:dpod} can be used analogously with the centered snapshots $\boldsymbol{\alpha}^c$ and the centered reduced basis $\bm{\mathsf{V}}^c$. The distance between the prediction $\boldsymbol{\tilde{\alpha}}_L(\boldsymbol{\mu})$ and the projection $\boldsymbol{\alpha}_L(\boldsymbol{\mu})$ is a measure for the closeness of the learned regression map and the observed map:
\begin{align}
	\begin{split}
	\delta_{\mathrm{REG}} \left(\boldsymbol{\mu}; \boldsymbol{\mathsf{V}}^c, \tilde{\boldsymbol{\pi}}^s \right) &= || \tilde{\boldsymbol{\alpha}}_L \left(\boldsymbol{\mu}\right) - \boldsymbol{\alpha}_L \left(\boldsymbol{\mu}\right)||_2 
	= || \tilde{\boldsymbol{\alpha}}^c_L \left(\boldsymbol{\mu}\right) - \boldsymbol{\alpha}^c_L \left(\boldsymbol{\mu}\right)||_2\\  
	& =  || \boldsymbol{\mathsf{V}}^c \tilde{\boldsymbol{y}}^c \left(\boldsymbol{\mu}\right) - \boldsymbol{\mathsf{V}}^c \boldsymbol{y}^c \left(\boldsymbol{\mu}\right)||_2 
	= ||\tilde{\boldsymbol{y}}^c \left(\boldsymbol{\mu}\right) - \boldsymbol{y}^c \left( \boldsymbol{\mu} \right) ||_2  \\ 
	& = || \boldsymbol{ \mathsf{ \Sigma } }^c \left( \tilde{ \boldsymbol{y} }^s \left( \boldsymbol{ \mu} \right) - \boldsymbol{y}^s \left(\boldsymbol{\mu}\right) \right) ||_2 
	= ||\boldsymbol{\mathsf{\Sigma}}^c \left(\tilde{\boldsymbol{\pi}}^s \left(\boldsymbol{\mu}^s\right) - \boldsymbol{\pi}^s \left(\boldsymbol{\mu}^s\right) \right) ||_2. 
	\end{split}
\end{align}
The total error is then the total distance between the true and predicted solutions:
\begin{align}
	\delta_{\mathrm{POD-REG}} \left( \boldsymbol{\mu}; \boldsymbol{\mathsf{V}}^c, \boldsymbol{\pi}^s \right) 
	= || \boldsymbol{\alpha} \left(\boldsymbol{\mu}\right) - \tilde{\boldsymbol{\alpha}}_L \left(\boldsymbol{\mu}\right)||_2 
	= || \boldsymbol{\alpha}^c \left(\boldsymbol{\mu}\right) - \tilde{\boldsymbol{\alpha}}^c_L \left(\boldsymbol{\mu}\right)||_2.
\end{align}
Eventually, any absolute error  $\delta_{\star}$ presented before can be turned into a standardized error $\boldsymbol{\varepsilon}_{\star}$, which is also the form we use in the evaluation of the ROM:
\begin{align}
	 \boldsymbol{\varepsilon}_{\star} \left(\mathbb{P}\right) \coloneqq \frac{ \delta_{\star} \left(\mathbb{P}\right)}{\sqrt{1/N_{tr} \sum_{\boldsymbol{\mu} \in \mathbb{P}_{\mathrm{tr}}} || \boldsymbol{\alpha} \left(\boldsymbol{\mu}\right) - \bar{\boldsymbol{\alpha}}||_2}}
\end{align}

%% file: input/sensitivityanalysis.tex
\subsection{Sensitivity Analysis}
\label{sec:SA}
We use a specific set of parameters for the inflow velocity and thus, the \gls{lvadlabel} speed algorithm. We want to investigate the influence of those parameters on washout, and how washout can be improved. To measure this influence we will use variance-based sensitivity analysis  \citep{saltelli2010} on two quantities, the washout (as defined in \cref{eq:washout}) \washout{} and concentration $c$. For \washout, we compute the \textit{total washout sensitivity} $\mathbb{S}_{i}^{\Delta c}$ for the $i$-th parameter $\mu_i$ as:
\begin{align}
\mathbb{S}_{i}^{\Delta c}= 1-\frac{\mathbb{V}[\mathbb{E}[\Delta c |\mu_i]] }{\mathbb{V}[\Delta c]} \label{eq:sa},
\end{align}
with the expected value $\mathbb{E}$, variance $\mathbb{V}$, and the output \washout. 
Sensitivity analysis will be performed with the \gls{romlabel} to decrease computational time of the high number of evaluations that are needed.
Furthermore, we evaluate the \textit{total concentration sensitivity} at each discrete location of the domain at the final simulation time of each snapshot as $\mathbb{S}_{i}^{c}$ for the i-th parameter $\mu_i$:
\begin{align}
\mathbb{S}_{i}^{c} = \mathbb{V}[c]-\mathbb{V}[\mathbb{E}[c|\mu_i]],
\end{align}
which is based on \cref{eq:sa}, but with the output $c$  and multiplied with the variance $\mathbb{V}[c]$ to highlight areas of high variance. We use the tool \textit{SALib} \citep{Herman2017}.

%% file: input/Stagnation_ROMS.tex
\section{Analyis of Reduced-Order Models for Concentration}
\label{sec:ROM-Washout}
We construct two \glspl{romlabel} for washout for two different inflow scenarios, the single pulse \gls{1plabel} and two consecutive pulses \gls{2plabel} with the parametrization as given in \Cref{eq:parametrization}.

The workflow to create the \gls{romlabel} is combined from the methods in \cref{sec:fom-fe} and in \cref{sec:ROM}:  $(i)$ generate the simulation data for training, testing, and validation,  $(ii)$ perform \gls{podlabel} on the snapshot matrix consisting of the final time step of all training set simulations, $(iii)$ perform regression on the reduced coefficients with \gls{gprlabel} and \gls{annlabel}. After creating the models, we check the model quality by computing the projection and prediction errors.

The two scenarios feature parameterized time-dependent boundary conditions for the \gls{lvadlabel}'s flow rate, prescribed as Dirichlet conditions. One is a single time-dependent sinusoidal pulse  (\gls{1plabel}), where the parameters are pulse period (\pulsePeriod{}) and the amplitude of the pulse ($A$). \pulsePeriod{} ranges between 2 and 5 seconds, $A$ is varied between 10 and 30 $\%$ of the baseline flow rate. The other scenario consists of two sinusoidal pulses (\gls{2plabel}), where the parameters are \pulsePeriod{} and the time between first and second pulse called pulse pause (\pulsePause{}), see \cref{fig:fluxdisc}. \pulsePeriod{} again ranges between 2 and 5 $\mathrm{s}$, \pulsePause{} is between 1 and 10 $\mathrm{s}$. The amplitude is kept constant at $120 \%$ of the baseline flow, for all parameters see \cref{tab:sim-params}. Apart from that, the same domain, boundary condition types, and simulation parameters are used. 

Both scenarios start from a steady state of constant LVAD baseline flow. All simulations of one scenario have the same total simulated time. In total, 144 samples have been simulated for \gls{1plabel} and for \gls{2plabel}, with 96 training samples, 24 test samples, and 24 validation samples.

\begin{figure}[h!]
    \begin{subfigure}{\textwidth}
        \centering
        \begin{subfigure}{0.32\textwidth}
            \centering

            \includegraphics[trim=0 0 0 25, clip, width=0.9\textwidth]{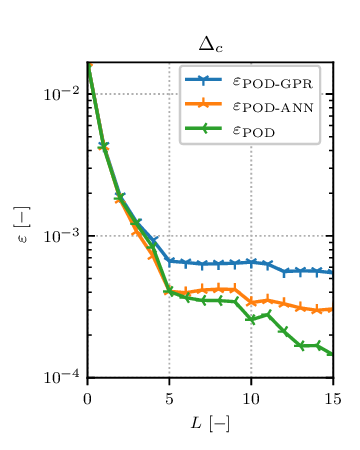}
            \vspace{5.75mm}
            \caption{\label{fig:errors_singlePulse_a}}   
        \end{subfigure}
        \begin{subfigure}{0.48\textwidth}
            \centering
            \includegraphics[width=0.9\textwidth]{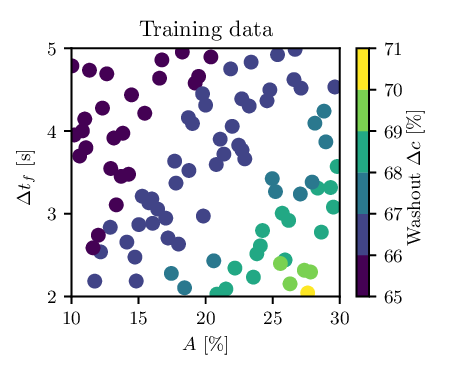}
            \vspace{-0.25mm}
            \caption{\label{fig:errors_singlePulse_b}}           
        \end{subfigure}
    \end{subfigure}
    \caption{(a) \gls{1plabel} \washout{} projection and prediction errors by regression method with respect to the number of basis functions used.  (b) Sampling of the training data sets from the parameter range colored by \washout{} values.}
    \label{fig:errors_singlePulse}
\end{figure}

For the \gls{1plabel} scenario, the $c$ projection error $\varepsilon_{\mathrm{POD}}$ and the $c$  prediction errors $\varepsilon_{\mathrm{POD-ANN}}$ and $\varepsilon_{\mathrm{POD-GPR}}$ for the \gls{annlabel} and \gls{gprlabel} regression are shown in \cref{fig:errors_singlePulse_a}. The projection error is the best result we can achieve with a given number of basis functions. We see that for the first 5 basis functions, regression by \gls{annlabel} closely follows the projection error, whereas \gls{gprlabel} produces a larger error that stagnates after 5 basis functions and only gets closer to the projection for more than 10 basis functions. \Gls{annlabel} stays closer to the projection error, but also stagnates between 5 and 10 basis functions. For more than 10 basis functions, \gls{annlabel} improves and is closer to the projection than \gls{gprlabel}. Overall, the average relative error in predicting \washout{} is below $10^{-3}$, when the range of \washout{} is $65\% \text{ to } 71\%$, as can be seen for the parameter samples drawn for the training data in \Cref{fig:errors_singlePulse_b}.

\begin{figure}[H]
    \begin{subfigure}{\textwidth}
        \centering
        \begin{subfigure}{0.32\textwidth}
            \centering
            \includegraphics[trim=0 0 0 25, clip, width=0.9\textwidth]{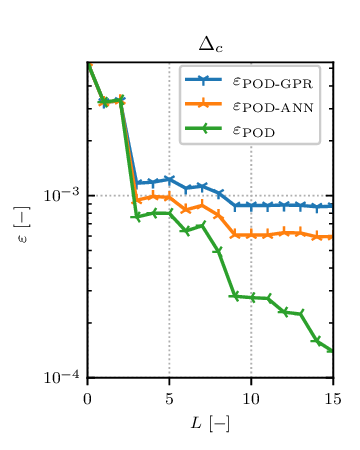}
            \vspace{5.75mm}
            \caption{\label{fig:errors_twoPulse_a}}   
        \end{subfigure}
        \begin{subfigure}{0.48\textwidth}
            \centering
            \includegraphics[width=0.9\textwidth]{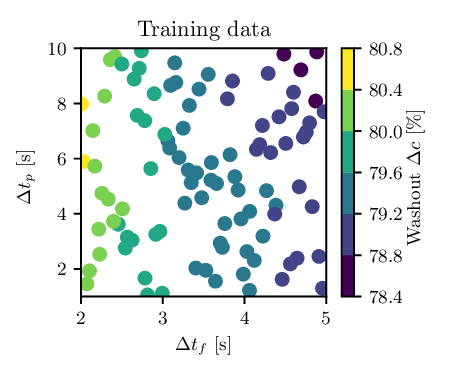}
            \vspace{-0.25mm}
            \caption{\label{fig:errors_twoPulse_b}}           
        \end{subfigure}
    \end{subfigure}
    \caption{(a) \gls{2plabel} \washout{} projection and prediction errors by regression method with respect to the number of basis functions used.  (b) Sampling of the training data sets from the parameter range colored by values of \washout{}.}
    \label{fig:errors_twoPulse}
\end{figure}

\begin{table}[h!]
		\begin{center}
		\caption{Simulation parameters for the two scenarios: \gls{1plabel} vs.  \gls{2plabel}.\label{tab:sim-params}}
			\begin{tabular}{ l c c l}
			\toprule
			\multicolumn{1}{l}{Quantity} & \gls{1plabel} & \gls{2plabel} & Unit \\
			\midrule
			Blood density $\rho$ &1.06 & 1.06& $  \mathrm{g/cm^3} $ \\
			Blood viscosity $\mu$ & 0.035 & 0.035 & $ \mathrm{s / cm^2} $ \\
			Amplitude $A$ & $[10,30]$ & $20$ & $\%$\\
			Pulse period \pulsePeriod & $[2,5]$ & $[2,5]$& s\\
                Pulse pause \pulsePause{} &  $\--$ & $[1,10]$ & s\\
			LVAD baseline flow $\dot{V}_{\mathrm{BL}}$ & 5 & 5 & $\mathrm{L/min}$\\
			Total simulation time $t_{\mathrm{end}}$ & 10 & 25 & s\\
			Time step $\Delta t$ & 0.1 & 0.1 & s\\
			\bottomrule
			\end{tabular}
		\end{center}
\end{table} 
\Cref{fig:errors_twoPulse_a} shows the errors and sample distribution for the \gls{2plabel} scenario. Both regression models follow the projection error closely up to the first two basis functions. With the third basis function, the projection error drops steeply, with \gls{annlabel} being closer to it than \gls{gprlabel}. Above 6 basis functions, both models decrease in error but stagnate beyond 10 basis functions, although the projection error still decreases. The relative error drops again below $10^{-3}$ when using more than 8 basis functions, with the \gls{annlabel} producing a lower prediction error than the \gls{gprlabel}. In this scenario, \washout{} lies between $78.4\%$ and $80.8\%$; see \cref{fig:errors_twoPulse_b}.

We find that for both scenarios, we can use the ROM to predict the concentration values with a relative error of less than $10^{-3}$, when we use 15 basis functions, and slightly more than $10^{-3}$ when we use only 5 basis functions. For the following usage of the \gls{romlabel}, we will use 15 basis functions, as the difference between using 5 or 15 basis functions is negligible with respect to computational cost, but the latter produces a slightly smaller error.

%% file: input/Results.tex
\section{Results} \label{sec:results}
The \glspl{romlabel} are used to investigate the effect of artificial pulsatility on the washout of the left ventricle under \gls{lvadlabel} support. 
Using the \gls{romlabel}  to predict \washout{} for the entire parameter range for both scenarios is faster than performing full-order simulations. Furthermore, we show how a predicted \gls{romlabel} solution - that is projected to the full-order space - is used to detect stagnation areas. 

For both scenarios, we quantify the influence of the parameters on the total \washout{} with sensitivity analysis. We also investigate the parameter sensitivity of the local concentration values on the full-order simulation domain, showing how stagnation zones can be influenced by varying the parameters.

\begin{figure}
\begin{centering}
        \centering
        \includegraphics[width=0.9\textwidth]{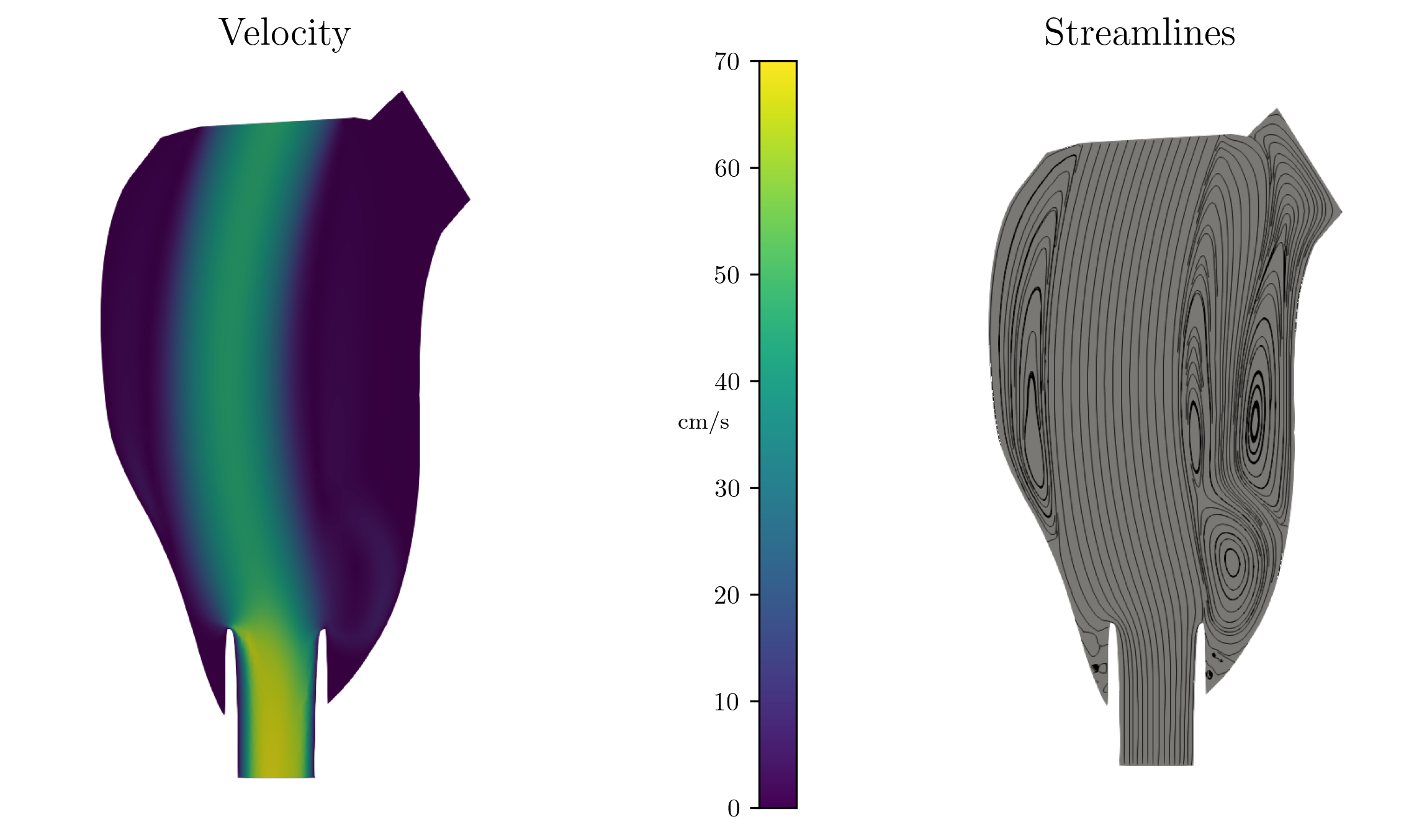}
        \vspace*{-0mm}
        \caption{Fluid velocity and streamlines at the end of the baseline flow simulation. \label{fig:baseline-plot}}
\end{centering}
\end{figure}

First, we present the flow field obtained by the baseline flow simulation of the left ventricle under \gls{lvadlabel} support in \cref{fig:baseline-plot}. The LVAD flow forms a strong main flow connecting the mitral valve on the upper left and the outflow. On the right, a recirculation zone forms that rotates counterclockwise from the apex back into the main flow. The left ventricular outflow tract around the aortic valve is characterized by very low velocities. On the other side of the main flow, a smaller recirculation zone forms because the flow is not completely directed against the ventricular wall, but enters the mitral valve at an angle. There are experimental studies presenting flow fields based on 2D imaging of mock loops, for example by \cite{WongStasis}, \cite{reider2017}, \cite{ortiz2021} and \cite{MayNewman}. Due to our simplified domain and boundary conditions, the comparison is only meaningful for experimental results from stages where the mitral valve is open and the aortic valve is closed. Although the images were acquired on transient fluid domains, the images also show a strong main flow from the mitral valve to the LVAD cannula, a breakup of the main vortex present in the non-LVAD left ventricle, and a stagnant region around the closed aortic valve. It should be noted that the recirculation zone is smaller compared to our results due to the angle of inflow. 

\color{black}
\subsection{Single Pulse ROM: Effect of Pulsatility on Washout for Single Pulse}
\begin{figure}
    \begin{subfigure}{\textwidth}
        \centering
        \begin{subfigure}{0.49\textwidth}
            \centering
            \includegraphics[width=0.9\textwidth]{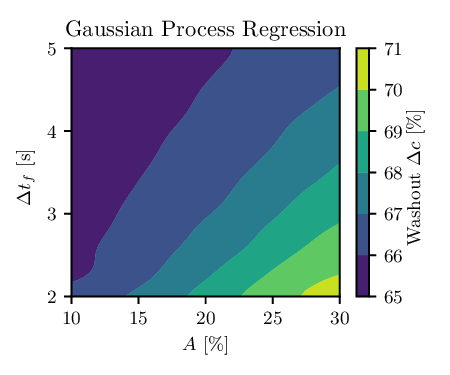}
        \end{subfigure}
        \begin{subfigure}{0.49\textwidth}
            \centering
            \includegraphics[width=0.9\textwidth]{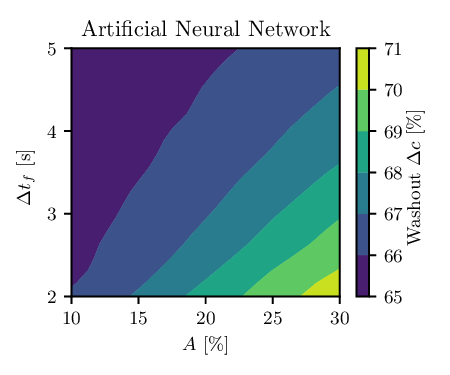}
        \end{subfigure}
    \end{subfigure}
    \caption{Comparison of the \gls{1plabel} \washout{} predicted by the regression based on \gls{gprlabel} and \gls{annlabel} over the parameter range spanned by \pulsePeriod{} and $A$.}
    \label{fig:models_singlePulse}
\end{figure}

The washout over the parameter space of the \gls{1plabel} scenario is presented in \cref{fig:models_singlePulse} for \gls{annlabel} and \gls{gprlabel}. It ranges from $65\% \textrm{~to~} 71 \%$. Both regression methods give a quantitatively very similar prediction for the parameter space. An increase of \washout{} is detected with increasing $A$ and decreasing \pulsePeriod{}. The maximum \washout{} is detected at the highest $A$ and lowest \pulsePeriod{}. 

For the non-pulsatile case, the washout is about $65 \%$. Non-pulsatile means that the baseline inlet flow rate is kept constant over the \gls{1plabel} total simulation $t_{\mathrm{end}}=10~\mathrm{s}$.
When choosing the best parameters versus the baseline flow, an increase in \washout{} of 9 $\%$ is reached in 10 $\mathrm{s}$. 

\begin{figure}
	\begin{subfigure}{\textwidth}
		\centering
		\begin{subfigure}{0.9\textwidth}
			\centering
			\includegraphics[width=0.9\textwidth]{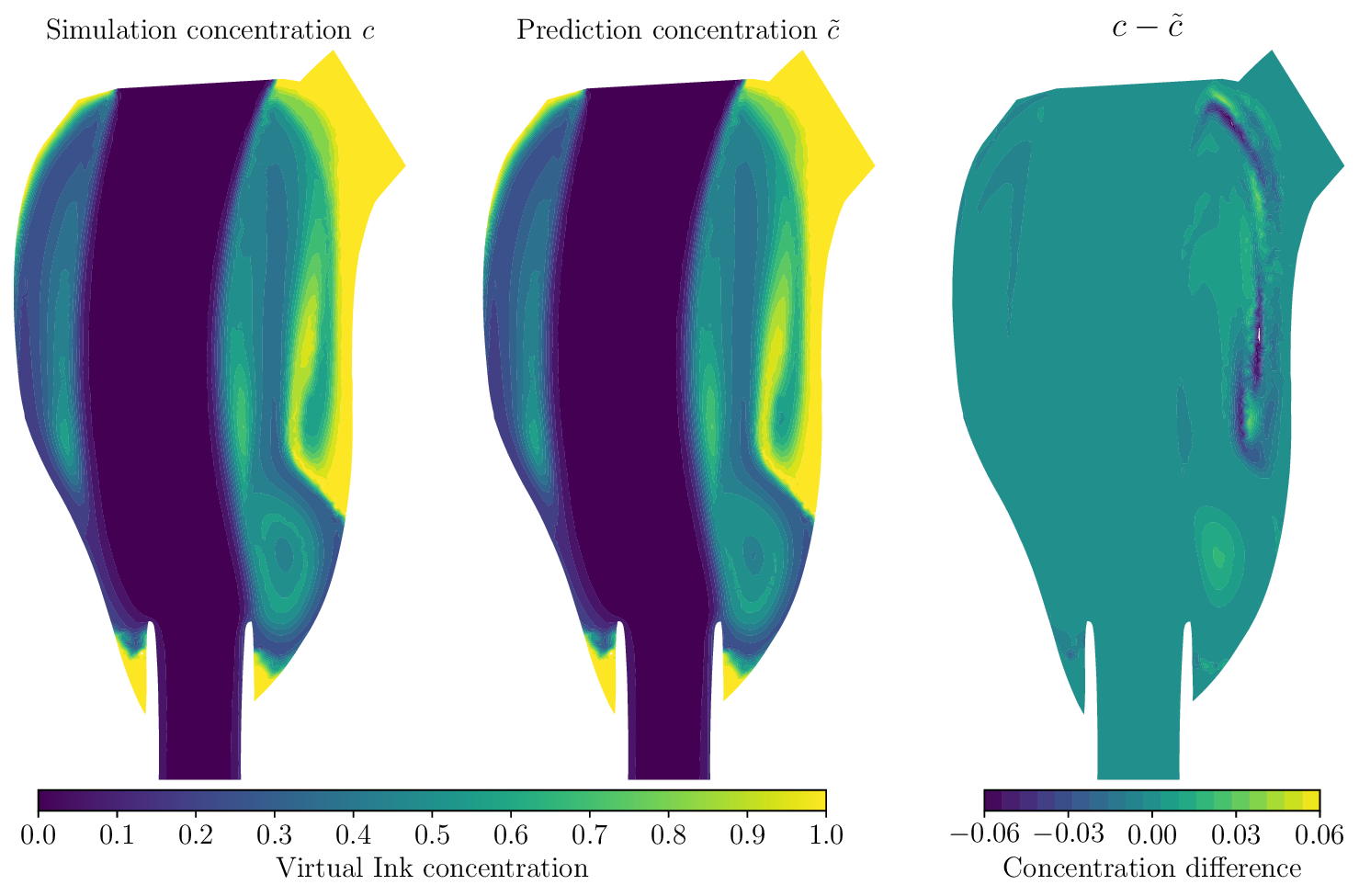}
			\caption{$L=5$}\label{fig:predic-vs-simul-b}
		\end{subfigure}
		\begin{subfigure}{0.9\textwidth}
			\centering
			\includegraphics[width=0.9\textwidth]{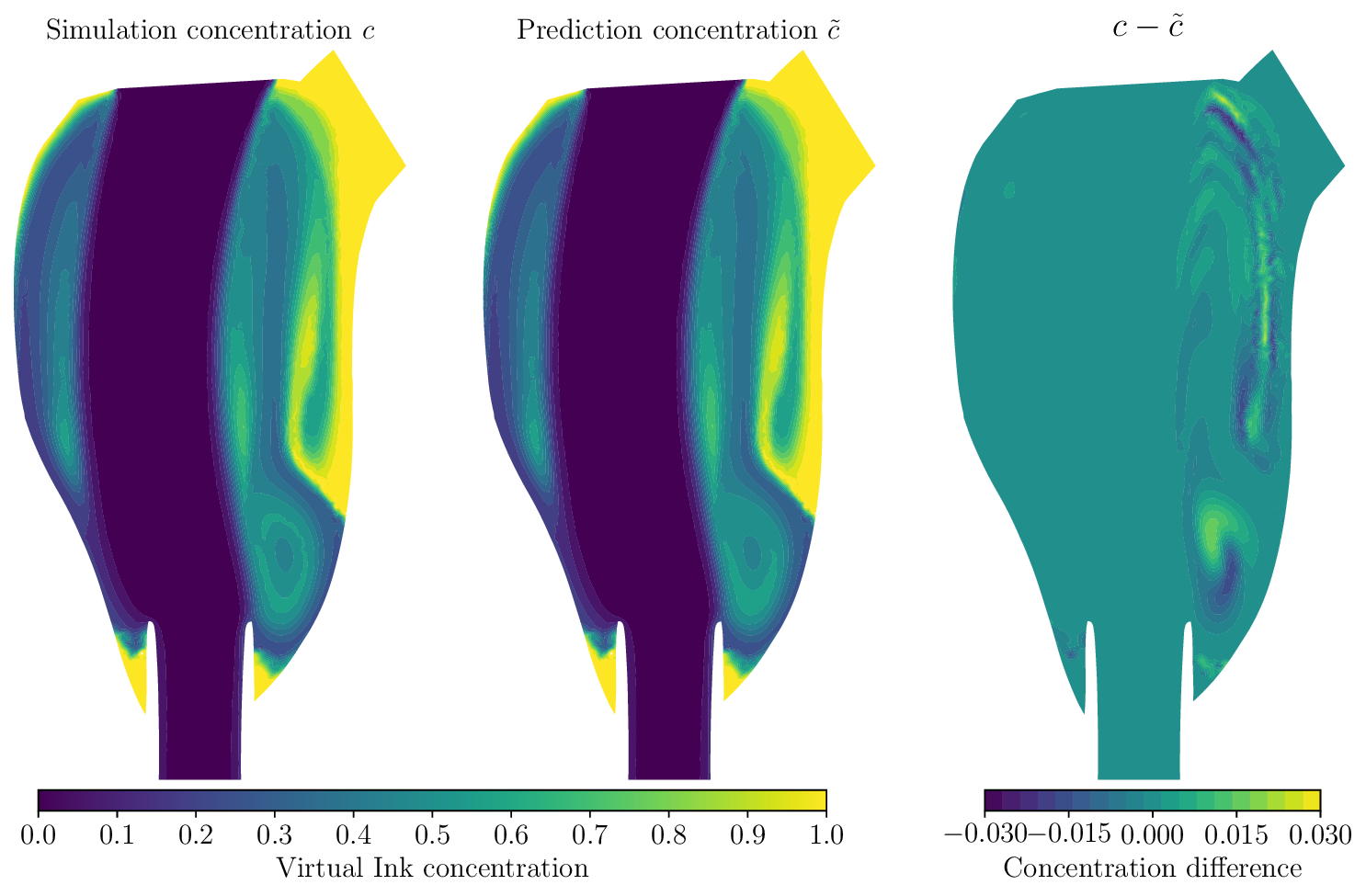}
			\caption{$L=15$}\label{fig:predic-vs-simul-a}
		\end{subfigure}
		
	\end{subfigure}
	\caption{Comparison of the results of a  \gls{1plabel} simulation (left) from the test data set to a prediction (middle) with \gls{gprlabel} of a parameter pair with high \washout{}  as presented in \cref{tab:predic-vs-simul} and the difference between both results (right) for 5 (upper row) and 15 (lower row) basis functions used for the reduced basis. \label{fig:predic-vs-simul}}
\end{figure}

The projection of a predicted \gls{romlabel} solution onto the full-order space $\tilde{c}$ is shown in comparison to a full-order solution $c$ from the \textit{validation} data set in \cref{fig:predic-vs-simul}. Within the range of concentration values, no visible difference is detectable. The rightmost plot shows the difference between both solutions. The difference between prediction and simulation is within $\pm 0.03$, equivalent to $6\%$ of the concentration range. The prediction accuracy is higher in the main flow and in the left recirculation area. In the right recirculation area, concentration values are under- and overestimated. The areas of stagnation with high concentrations around the aortic valve and the cannula are represented similarly in \gls{romlabel} and full-order simulation. 

From the evaluation in \cref{fig:models_singlePulse} we have seen that for more than 5 basis functions, we get closer to the \gls{podlabel}. This can be seen in \cref{fig:predic-vs-simul-b} and \cref{fig:predic-vs-simul-a}, which show the difference in using 15 vs. 5 basis functions. The local error in concentration prediction ranges between $\pm0.06$ for $L=5$, twice as high as for $L=15$. It is apparent that the first 5 basis functions cannot capture the complex influence of the input parameters on the flow field, especially in the recirculation zones. 

\Cref{tab:predic-vs-simul} presents the \washout{} values for two samples of the \gls{1plabel} scenario. The difference between predicted and simulated \washout{} is $1\cdot 10^{-4}$ for the high washout case, and $8\cdot 10^{-5}$ for the low washout case. 

\subsection{Single Pulse ROM: Sensitivity Analysis of Pulse Period and Amplitude}

In \cref{tab:sensitivities_washout} total sensitivity of \washout{} with respect to the parameters is given. Both parameters have a significant influence and $\mathbb{S}^{\Delta_c}_A$ is higher than  $\mathbb{S}^{\Delta_c}_{\Delta t_{f}}$. 

Performing sensitivity analysis on the concentration field yields the absolute concentration sensitivities for $A$ and \pulsePeriod{}; see \cref{fig:single-pulse-sa}. We evaluate the sensitivities at each discrete location. The plot shows how strong the concentration at each discrete location reacts to changes in the input parameters. Changing $A$ leads to a change in concentration in the lower part of the right recirculation area, whereas a change in \pulsePeriod{} has an effect on the upper parts of the recirculation area, closer to the aortic valve. A shorter pulse can transport new blood deeper into this area at the same amplitude.

\subsection{Two Pulse ROM: Effect of Pulsatility on Washout for Two Pulses}
For \gls{2plabel}, we present washout over the parameter ranges for \pulsePeriod{} and \pulsePause{} in \cref{fig:models_twoPulse}. It ranges between $78.4\%$ and $80.8\%$.  Washout and the total simulation time are higher in \gls{2plabel}. Both regression methods show the maximum washout for medium to high \pulsePause{} around $6-8~\mathrm{s}~$ (\gls{gprlabel}) and $4-10~\mathrm{s}~$(\gls{annlabel}) and low $\Delta t_f=2~\mathrm{s}$. For values between maximum and minimum, the washout is independent of  \pulsePause{}, but decreases with increasing \pulsePeriod{}. 

Non-pulsatile flow for the length of \gls{2plabel} scenario results in \washout{} $=76.9\%$. The increase in \washout{} for the pulsatile case over the non-pulsatile case is 5 $\%$ over the time period of $25~\mathrm{s}$.

\subsection{Two Pulse ROM: Sensitivity Analysis of Pulse Period and Pulse Pause}
\begin{figure}
    \begin{subfigure}{\textwidth}
        \centering
        \begin{subfigure}{0.49\textwidth}
            \centering
            \includegraphics[width=0.9\textwidth]{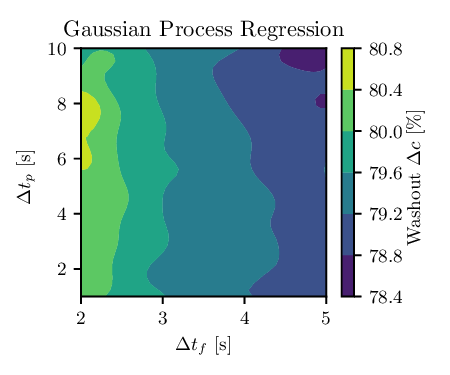}
        \end{subfigure}
        \begin{subfigure}{0.49\textwidth}
            \centering
            \includegraphics[width=0.9\textwidth]{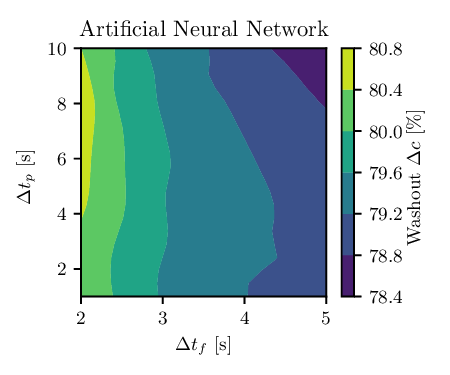}
        \end{subfigure}
    \end{subfigure}
    \caption{Comparison of the \gls{2plabel} \washout{} predicted by regression based on \gls{gprlabel} and \gls{annlabel} over the parameter range spanned by \pulsePause{} and \pulsePeriod{}.}
    \label{fig:models_twoPulse}
\end{figure}

\begin{figure}
\begin{centering}
    \begin{subfigure}{0.7\textwidth}
        \centering
        \includegraphics[width=0.9\textwidth]{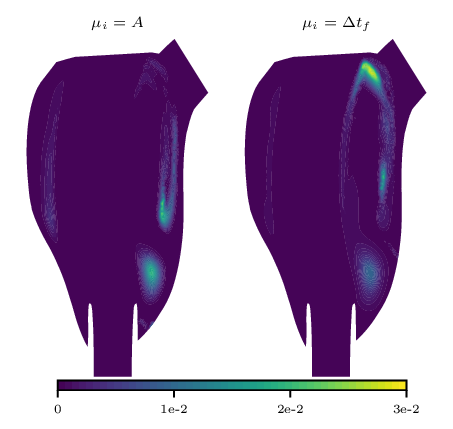}
        \vspace*{-0mm}
        \caption{Single pulse scenario (\gls{1plabel}). \label{fig:single-pulse-sa}}
     \end{subfigure}
    \begin{subfigure}{0.7\textwidth}
        \centering
        \includegraphics[width=0.9\textwidth]{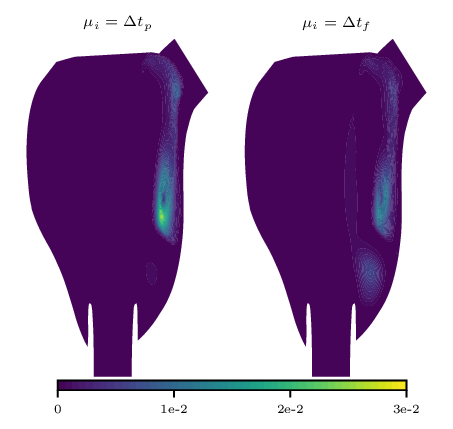}
        \vspace*{-0mm}
        \caption{Two pulse scenario (\gls{2plabel}). \label{fig:two-pulse-sa}}
     \end{subfigure}
      \vspace*{-4mm}
      \caption{Absolute concentration sensitivities. \label{fig:sa}}
\end{centering}
\end{figure}

\Cref{tab:sensitivities_washout} contains the sensitivities with respect to the parameters in the \gls{2plabel} scenario. The total washout sensitivity of \pulsePeriod{} is close to $1$, whereas the value for \pulsePause{} is close to zero. This supports the findings of the dominance of \pulsePeriod{} for influencing \washout{}, see the nearly vertical contour lines in \cref{fig:models_twoPulse}. 
The sensitivity of the concentration $c$ with respect to \pulsePeriod{} and \pulsePause{} is presented in \cref{fig:two-pulse-sa}. The areas that are sensitive to changes in \pulsePeriod{} spread across the entire right recirculation zone with emphasis on the vortical structure in the middle part. No influence around the cannula or in the left part of the domain is detected. Compared to \gls{1plabel}, the \pulsePeriod{} sensitivity is less pronounced but visible up to the aortic valve. This could be also due to the decreased range of concentration values, as also detected in \cref{fig:models_twoPulse}. The influence of \pulsePause{} is focused around the right recirculation zone and mostly in the middle and upper part. Local values are higher than for \pulsePeriod{} and more distinct in the aortic valve region. 

\begin{table}
\begin{center}
    \small
\caption{Comparison of the washout $\Delta c$ for two samples from the test set for the \gls{1plabel} scenario.  \label{tab:predic-vs-simul}}
\begin{tabular}{r r r}
\toprule
\addlinespace[2pt]
 & High \washout & Low \washout \\
\addlinespace[2pt]
\midrule
A $[\%]$  &  28.7 & 14.1\\
$\Delta t_f ~[\mathrm{s}]$ &2.46 &4.58 \\
$\Delta c_{S}$ & 0.69745 & 0.65741\\ 
$\Delta c_{L}$  & 0.69735 & 0.65749\\
$\Delta c_{S}$ - $\Delta c_{L}$ &$1\cdot 10^{-4}$  & $8\cdot 10^{-5}$\\
\addlinespace[2pt]
\bottomrule
\end{tabular}
\end{center}
\vspace{-5mm}
\end{table}
\begin{table}
    \centering
    \small
     \caption{Total washout sensitivity with respect to the parameters of the two scenarios.}
    \label{tab:sensitivities_washout}
    \begin{tabular}{lcccc}
	\toprule         
         &
        \multicolumn{2}{c}{\gls{1plabel}} & 
        \multicolumn{2}{c}{\gls{2plabel}} \\
        \addlinespace[5pt]
		\cmidrule(lr){2-3} \cmidrule(lr){4-5}        
        & 
        $A$ & \pulsePeriod{} &
        \pulsePeriod{} & \pulsePause{}\\
		\cmidrule(lr){2-3} \cmidrule(lr){4-5}
        Total washout sensitivity $\mathbb{S}_{i}^{\Delta {c}}$  & 0.613 & 0.453 & 0.971 & 0.055 \\
        Confidence interval size  & 0.083 & 0.059 & 0.100 & 0.008 \\
         \bottomrule
    \end{tabular}
\end{table}

\subsection{Computational Time}
The computational cost of running the 2D simulations, training the regression models, and evaluating the ROMs for sensitivity analysis is shown in \cref{tab:timings}. The estimated timings for training the regression models refer to the cost for one particular regression model, as they were similar for GPR and ANN. The same is true for a single evaluation of the ROM, so the table reflects the total time taken to perform the sensitivity analysis for the \gls{1plabel} and \gls{2plabel}, respectively. Once the ROM is created, the evaluation of the ROM for an unseen parameter sample can be obtained instantly. To calculate the speed-up we can achieve by using the ROMs for sensitivity analysis, instead of generating these data points by simulation, we need to compare all the costs involved. Training, testing, and validating each scenario requires 144 simulations, with average runtimes of 11734 s and 13906 s. These wall clock times were obtained using two CPU cores per simulation, so the total core time is $ 3.28 \cross 10^{6}$ s and $ 4 \cross 10^{6}$ s,respectively. We assume that the same number of full-order simulations would have been required to perform the sensitivity analysis, and note that the generation of the training, test, and validation data sets is the major contributor to the cost of the ROMs. The cost for training the ROMs and evaluating them for sensitivity analysis is negligible in comparison. 

Then the speed-up S is the ratio of the number of evaluations performed for the sensitivity analysis to the number of data points required to generate the ROMs. Thus, the speed-up can be calculated as $S = \frac{3072}{144} \approx 21$.

\begin{table}
\footnotesize
\centering
\caption{Run-time comparison of the full-order simulation, the training of one regression model, and the evaluation of the ROM to create the sensitivity analysis. The number of ROM evaluations according to the SALib documentation is $N_\textrm{eval} = N (2 D + 2)$, with the number of parameters $D=2$ and the number of evaluations $N=512$. The presented wall-clock times for the \gls{1plabel} and \gls{2plabel} scenario are average run times of the training simulations.}\label{tab:timings}
\begin{tabular}{p{2.5cm} r r r r}
\toprule
    Task & Wall-clock time & Times performed & Cores &  Total core time \\
    \midrule
     Full-order simulation \gls{1plabel} &  11734 s & 144 & 2 &  3275712 s\\
     Full-order simulation \gls{2plabel} &  13906 s & 144 & 2 &  4004928 s\\
     Regression model training & $<$ 900 s & 1 & 1 & $<$ 900 s \\
     ROM evaluation for sensitivity analysis (\gls{1plabel}) & [-] & 3072 & 1 & $<$ 60 s  \\
     \bottomrule
\end{tabular}
\end{table}

\color{black}

%% file: input/Discussion.tex
\section{Discussion}
\label{sec:discussion}
The presented combination of full-order simulation, model order reduction, and sensitivity analysis has shown a significant speed-up compared to using simulations alone. The speed-up is driven by the drastically reduced cost of simply evaluating the \gls{romlabel} for sensitivity analysis instead of running more simulations. The proposed method for \gls{romlabel} is non-intrusive, so it does not affect the simulation itself. The projection of the predicted solutions allows to evaluate the sensitivity with respect to local quantities. The presented workflow can be applied to more clinically accurate applications and simulations.

In our simplified simulation scenarios, the \glspl{romlabel} are able to predict \washout{} with an error of $\mathcal{O} ( 10^{-4})$. They are able to identify stagnation areas at the aortic valve and around the connection between LVAD cannula and left ventricle apex by reconstructing the \gls{romlabel} solution to the full-order space. Those regions are also reported in literature, see \cite{reider2017}. Using the predictions of the \gls{romlabel} instead of the full-order simulation shows no visible difference in the concentration values (see  \cref{fig:predic-vs-simul}), but evaluation of the \gls{romlabel} takes seconds, compared to many CPU hours for the full-order simulation. The integral quantity \washout{} is less sensitive to the numerical differences in concentration prediction than the scalar concentration $c$, which could be because of the symmetries in $c_S - c_L$, as visible in \cref{fig:predic-vs-simul}.


In this study, adding pulsatility to the \gls{lvadlabel} flow rate improves washout in both scenarios. In both cases, pulsatile flow leads to higher or at least equivalent washout values than non-pulsatile flow. Having no pulsatility is equal to a zero amplitude and no \pulsePeriod{}. In case of the \gls{1plabel}, the lower $A$ will decrease \washout{} and there is no gradient in flow rate. In case of \gls{2plabel}, there is also no gradient in flow rate to promote washout, especially around the aortic valve.

In the \gls{1plabel} scenario, short, high amplitude pulses wash out the ventricle better than longer pulses or pulses equivalent in length but lower in amplitude. The nearly linear increase in \washout{} with the two parameters implies a proportional relationship between \washout{} and the product $A \cdot \Delta t_{f}$. The stagnation zones are prominent around the cannula at the apex and at the aortic valve in both scenarios. The sensitivity analysis shows that $A$ has a high influence on the lower right recirculation zone, whereas \pulsePeriod{} shows a higher sensitivity close to the aortic valve in the upper right recirculation zone. 

In the \gls{2plabel} scenario, the time that passes between two pulses does not seem to influence the washout very much, at least for the time range presented in this study. Again, the length of a pulse has a higher influence on the concentration field. This observation is common for both scenarios. The shorter pulse has a faster variation in time if the amplitude is fixed. We suppose that the higher gradient of flow rate allows the incoming blood to reach deeper into the recirculation zone around the aortic valve and wash out the left ventricle better. This would also explain the sensitivity of the concentration field with respect to \pulsePeriod{} in that region, see \cref{fig:sa}. The sensitivity analysis shows an influence of \pulsePause{} in the middle of the right recirculation zone, but \pulsePeriod{} has more influence in the lower recirculation area. Error analysis shows that even when using a limited number of basis functions, the \glspl{romlabel} produce a relative error of $10^{-3}$ for the reduced coefficients. 

Our simulations produce flow velocities in the left ventricle with LVAD and closed aortic valve that are qualitatively similar to experimental images \citep{WongStasis,MayNewman,reider2017,ortiz2021}. Short, high-amplitude pulses are now available in commercial LVADs, such as the Heartmate3 with an artificial pulse. Such a pulse increases the kinetic energy in the left ventricle \citep{maurer2022}. This could explain our findings of highest washout for short, intense pulses in the \gls{1plabel} scenario.

The presented simulation setup of the left ventricle with LVAD is a test case and therefore simplified. We did not use a transient, three-dimensional domain and no turbulence model was investigated. A realistic movement of the left ventricular wall is currently not included. The boundary conditions of the flow simulation were not determined by a cardiovascular model, so it was not possible to include intermittent aortic valve opening. Washout was only tracked at the last time step and not as a function of time. 

The method of combining a non-intrusive \gls{romlabel} and sensitivity analysis can be used with other parameterized problems. We plan to use a realistic, transient 3D domain from imaging. A turbulence model should be included for the flow in the transient regime. This would make the results of the flow simulation more diverse and thus could have an influence on the washout as an advective quantity. The boundary conditions will be determined with a 0D lumped-parameter network to allow for intermittent aortic valve opening. This will have an effect on the washout, especially in the regions close to the aortic valve. Full-order, 3D simulations would be much more expensive. If the cost of creating and evaluating a \gls{romlabel} of such a 3D case does not increase proportionally to the simulation cost of 3D simulations, using the proposed workflow will lead to comparable speed-ups.
Similar to this study, other parameters such as cannula insertion length, cannula angle, and cannula position could be investigated with respect to washout or other hemodynamics. Synchronization of LVAD velocity modulation with native left ventricular motion could also be investigated as a parametrized problem.

%% file: input/Conclusion.tex
\section{Conclusion}
\label{sec:conclusion}

We have presented two ROMs with POD of a snapshot matrix based on full-order simulations of an idealized ventricle with LVAD discretized with FEM for two scenarios of transient LVAD rotational speed. We have quantified the washout in the left ventricle based on the final concentration of freshly injected blood obtained with the virtual ink method.
The reduced computational time needed for evaluating the \glspl{romlabel} allows for a lot of queries to investigate the sensitivity of $c$ and \washout{} with regard to the parameters in both scenarios. The speedup is driven by the nearly instantaneous evaluation of the ROM for sensitivity analysis compared to using full order simulations. It then depends on how many queries the sensitivity analysis requires.

The \glspl{romlabel} have been compared and verified by computing the prediction error on the validation data set. A low number of basis functions is sufficient to approximate the reduced coefficients needed to predict a solution within a relative error of $\mathcal{O} (1 \cdot 10^{-3})$. The prediction of the concentration $c$ at each location of the 2D left ventricle is prone to over- and underestimations within a range of $\pm 0.06\%$. The small errors in prediction result in an accurate approximation of \washout{}.

We showed the possibility of plotting the sensitivity of $c$ at each discrete location to investigate the influence of changing parameters on local stagnation zones. With a reasonable number of simulations, a full picture of the influence of the flow rate pulsatility on washout in the left ventricle is given. In the future,  \gls{romslabel} for washout in realistic, 3D simulations could be obtained with the presented workflow. 
Our results confirm the benefit of including pulsatility in \gls{lvadlabel} flow rate to improve washout in the scope of a numerical study. For two scenarios, pulsatility has increased the washout by up to 9 $\%$ when using the optimal parameters compared to the non-pulsatile baseline flow. 

Stagnation zones have been identified close to the aortic valve and around the \gls{lvadlabel} cannula.
The two scenarios show that in the single pulse case short amplitude pulses wash out the ventricle best. For two pulses, the time between two subsequent pulses plays a minor role, whereas again short pulses are favorable over longer ones to improve washout. We have shown that a pulsatile LVAD pump flow rate washes out the left ventricle better than a non-pulsatile flow rate. 

The presented workflow gives insight into the effects of pulsatility parameters, global stagnation values and local stagnation zones and allows a systematic approach to find optimal parameters to reduce stagnation.

\section*{Acknowledgements}
The authors gratefully acknowledge the computing time granted by the JARA Vergabegremium and provided on the JARA Partition part of the supercomputer CLAIX at RWTH Aachen University.

The authors gratefully acknowledge the computing time provided to them on the high-performance computer Lichtenberg at the NHR Centers NHR4CES at TU Darmstadt. This is funded by the Federal Ministry of Education and Research, and the state governments participating on the basis of the resolutions of the GWK for national high performance computing at universities (www.nhr-verein.de/unsere-partner).

The research project was funded by the Federal Ministry of Economic Affairs and Climate Action under the  “Central Innovation Programme for small and medium-sized enterprises”(ZIM). The authors would like to thank for the support.

This work was funded by the Deutsche Forschungsgemeinschaft (DFG, German Research Foundation) through grant 333849990/GRK2379 (IRTG Modern Inverse Problems).